\documentclass[11pt]{article}

\def\be{\begin{equation}}
\def\ee{\end{equation}}
\def\ba{\begin{eqnarray}}
\def\ea{\end{eqnarray}}
\def\la{\langle}
\def\ra{\rangle}
\def\a{\alpha}

\def\h{\hskip 1cm}

\usepackage{graphicx}
\begin{document}
\begin{titlepage}
\vspace{4cm}
\begin{center}{\Large \bf Equi-entangled bases in arbitrary dimensions} \\
\vspace{1cm}V. Karimipour\footnote{Corresponding author,
email:vahid@sharif.edu},
\hspace{0.5cm} L. Memarzadeh \footnote{email:laleh@mehr.sharif.edu}\\
\vspace{1cm} Department of Physics, Sharif University of Technology,\\
P.O. Box 11365-9161,\\ Tehran, Iran
\end{center}
\vskip 3cm
\begin{abstract}
For the space of two identical systems of arbitrary dimensions, we
introduce a continuous family of bases with the following
properties: i) the bases are orthonormal, ii) in each basis, all
the states have the same values of entanglement, and iii) they
continuously interpolate between the product basis and the
maximally entangled basis. The states thus constructed may find
applications in many areas related to quantum information science
including quantum cryptography, optimal Bell tests and
investigation of enhancement of channel capacity due to
entanglement.
\end{abstract}
\end{titlepage}
\section{Introduction}
In quantum computation and quantum information theory
\cite{nielsen,machrev} and many of the related fields the well known
Bell basis finds numerous applications. This basis consists of four
ortho-normal maximally entangled states of two qubits, denoted by:

\begin{eqnarray}
  |\phi^{\pm}\ra &=& \frac{1}{\sqrt{2}}(|0,0\ra \pm |1,1\ra), \\
  |\psi^{\pm}\ra &=& \frac{1}{\sqrt{2}}(|0,1\ra \pm |1,0\ra).
\end{eqnarray}

The analogs of these states have been defined for $d-$ dimensional
systems, or qudits in the following form
\begin{equation}\label{gbell1}
    |\psi_{m,n}\ra:=
    \frac{1}{\sqrt{d}}\sum_{j=0}^{d-1}\xi^{jn}|j,j+m\ra, \ \ \  m,n=0\cdots d-1.
\end{equation}
where $\xi := e^{\frac{2i\pi}{d}}$ and $\{|0\ra, \cdots |d-1\ra\}$
is an orthonormal basis for the space of one qudit. The states
(\ref{gbell1}) are maximally entangled and are mutually orthogonal.
These states have found applications in almost every generalization
of quantum algorithms and protocols and investigations of
entanglement in higher dimensional
systems \cite{alb1,alb2,fiur,kar,woo, mint,gitt,brauns,vah}. \\

The present paper poses a simple
question. \\

Can we construct a basis for the space of two qudits which has the
following properties: \\

1- It is an orthogonal basis,\\

2- all the states have the same value of entanglement, and\\

3- by varying a set of parameters, the basis while retaining the
above two properties, continuously changes from a product basis to a
maximally entangled basis?\\

We call such a basis an equi-entangled orthogonal basis or an
interpolating basis. This question has a simple and well-known
answer for qubits, namely
\begin{eqnarray}
  |\phi^+(\theta)\ra &=& \cos \theta |0,0\ra + \sin \theta |1,1\ra, \\
  |\phi^-(\theta)\ra &=& \sin \theta |0,0\ra - \cos \theta |1,1\ra, \\
  |\psi^+(\theta)\ra &=& \cos \theta |0,1\ra + \sin \theta |1,0\ra, \\
  |\psi^-(\theta)\ra &=& \sin \theta |0,1\ra - \cos \theta |1,0\ra.
\end{eqnarray}
This basis continuously connects a product basis (for $\theta = 0 $)
to a maximally entangled basis (for $\theta = \frac{\pi}{4}$). \\
We should stress that it is quite trivial to construct one single
state which interpolates between a product state and a maximally
entangled state. What we want is an orthonormal basis which connects
a product basis to a maximally entangled basis. Moreover we require
that all along the way each of the basis states has the same value
of entanglement. \\
To our knowledge such a construction has not been generalized to
higher dimensions and it is our aim in this paper to obtain such an
interpolating basis in higher dimensions. This construction turns
out to be quite non-trivial, since it requires the solution of
simultaneous non-linear equations the number of
which increase with dimension.\\

Before proceeding we would like to point out some probable
applications of these kinds of states.\\

In an interesting recent development it has been shown that
optimal Bell tests do not require maximally entangled states. In
fact as stated in \cite{gis}, it has often been assumed that
generalized Bell states of the form (\ref{gbell1}) represent the
most nonlocal quantum states while in \cite{gis}, it is shown
that optimal states for tests of non-locality are not maximally
entangled except in the case of qubits indicating that the role
of dimension is more subtle than previously supposed. Therefore
having a set of mutually orthogonal states, all with the same
value of entanglement, may facilitate the search for optimal
states that test non-locality in arbitrary dimensions. In effect
having such an equi-entangled basis generalizes the concept of
Bell state measurement so that the measurement projects the
measured states into mutually orthogonal states with a
prescribed and tunable value of entanglement. \\

There is an important question in quantum information theory: is
the classical capacity of quantum channels still additive if we
allow encoding of input data into entangled states? Although
there are clues that in channels without memory \cite{faoro,king}
(where there is no correlation between two consecutive uses of
the channel), entanglement does not enhance the capacity, for
channels with memory there are clues that this is not the case
\cite{mach1,mach2}. Depending on the value of correlations and
the type of channel, one may want to encode the data (the
alphabet or the strings) into mutually orthogonal and hence
distinguishable quantum states with a prescribed and not
necessarily maximal value of entanglement. The analysis of such a
channel will then be made much easier if all the states have the
same value of entanglement. In other words such states make it
possible to have mixed states of arbitrary entanglement by taking
$$\rho = \sum_i{P_i|\psi_i\ra\la \psi_i|},$$ where $\{|\psi_i\ra\}$
is an equi-entangled
orthonormal basis.\\

Other areas in which such states may find applications are quantum
cryptography, quantum dense coding and quantum teleportation
\cite{nielsen}. In all these areas Bell states and Bell state
measurements play an essential role and all of them have been
generalized to arbitrary dimensions. A generalized Bell state
measurement in which the degree of entanglement of the output states
can be tuned to any arbitrary value, may someday find
applications in these areas.\\

Presumably these are not the only applications that one may envisage
for these kinds of states.\\

The structure of this paper is as follows: In section \ref{general}
we present the general construction of these states in arbitrary
dimensions and obtain the system of quadratic equations which should
be solved explicitly in order to find the concrete form of the
states. In sections \ref{three} and \ref{four} we present explicit
solutions of the equations in three and four dimensions. Finally in
section \ref{higher} we obtain the general solution of the quadratic
system of equations in arbitrary dimensions.

\section{A method of construction }\label{general}
Here we present a method for construction of equi-entangled bases.
Consider the single state
\begin{equation}\label{phi00}
    |\psi_{0,0}\ra:=\sum_{i=0}^{d-1}a_i|i,i\ra,
\end{equation}
and the unitary shift operator
\begin{equation}\label{}
    S:=\sum_{i=0}^{d-1} |i+1\ra\la i|.
\end{equation}
For $m,n=0,1,\cdots d-1$, define $d^2$ states of the form

\begin{equation}\label{}
    |\psi_{m,n}\ra = S^{m}\otimes S^{m+n} |\psi_{00}\ra = \sum_{i=0}^{d-1} a_i |i+m, i+m+n\ra.
\end{equation}

All these states have the same value of entanglement as the state
$|\psi_{0,0}\ra$, since all of them are obtained from
$|\psi_{0,0}\ra$ by bi-local unitary operators. The value of
entanglement is obtained by using von Neumann entropy
\begin{equation}\label{ent}
    E(|\psi_{m,n}\ra)=
    E(|\psi_{00}\ra)=-\sum_{i=0}^{d-1}|a_i|^2\log_d |a_i|^2.
\end{equation}

Note that we have taken the logarithms in base $d$ so that a
maximally entangled state has a value of entanglement equal to unity
in all dimensions.\\

We now demand that they be ortho-normal. This leads to the following
set of equations:

\begin{equation}\label{ortho}
    \sum_{i=0}^{d-1}\overline{a_i}a_{i+m}=\delta_{m,0}\h m=0,\cdots
    d-1.
\end{equation}

This is a set of $d$ quadratic equations for the complex
coefficients $a_i$. The solution of this set of equations gives the
final form of the state $|\psi_{00}\ra$ and hence all the state
vectors of the basis.\\

A natural question arises as to whether this is the only possible
construction for obtaining such bases. We have tired many other
constructions and have found that the ansatz presented in this
section gives the most elegant or even inevitable solution. In fact
the most general way of constructing $d^2$ equi-entangled states
from $|\psi_{00}\ra$ is to define the states as
$|\psi_{mn}\ra=U_m\otimes V_n|\psi_{00}\ra$ where $\{U_m\}$ and
$\{V_n\}$ are a set of $2d$ unitary operators. However orthogonality
puts a stringent requirement on these operators in the form
\begin{equation}\label{form}
    \la \psi_{00}|U_m^{\dagger}U_p\otimes
    V_n^{\dagger}V_q|\psi_{00}\ra = \delta_{m,p}\delta_{n,q}.
\end{equation}
One can simplify the situation by taking $U_m=U^m$ and $V_n=V^n$,
hence searching for two unitary operators $U$ and $V$ with the
property that $\la \psi_{00}|U^m\otimes
V^n|\psi_{00}\ra=\delta_{m,0}\delta_{n,0}$. Calculations then show
that the choice $U=V=S$ where $S$ is the shift operator defined
above gives the most elegant solution. Despite this we refrain to
claim that our construction is a general one. \\

Before proceeding to the general solution, for concreteness, we
consider the cases of three and four dimensional spaces separately.

\section{Equi-entangled basis for three dimensional
spaces (qutrits)}\label{three}
 \subsection{Case a: Real vector space}
\begin{figure}[t]
 \centering
   \includegraphics[width=12cm,height=8cm]{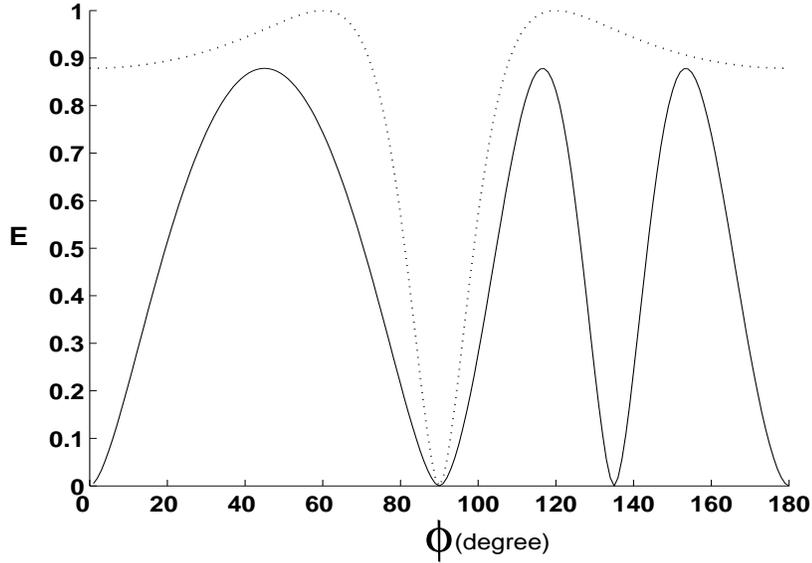}
   \caption{a)The entanglement of the basis (\ref{3r}) as a function of $\phi$ (in degree)(-----).
            b)The entanglement of the basis (\ref{3c}) as a function of $\phi$ (in degree)(.....). The entanglement is dimensionless.}
   \label{E3a}
\end{figure}

If we assume that the basis vector $|\psi_{00}\ra$ is real, all the
other vectors will be real and we can construct a basis for a real
vector space. In this case equations (\ref{ortho}) reduce to the
following two equations
\begin{eqnarray}\label{3}
    a_0^2+a_1^2+a_2^2 &=&1\cr &&\cr
    a_0a_1+a_1a_2+a_2a_0&=&0.
\end{eqnarray}
Parameterizing the variables in the first equation by polar
coordinates

\begin{equation}\label{par}
    a_0=\sin \theta \cos \phi, \ \ \ a_1=\sin\theta \sin \phi, \ \ \
    a_2=\cos \theta,
\end{equation}
and inserting these into the second equation we obtain a relation
between the parameters, namely
\begin{equation}\label{par2}
\tan \theta = -\frac{\sin \phi + \cos \phi}{\sin \phi \cos \phi}.
\end{equation}
Taking this relation into account we find the final form of the
coefficients:

\begin{eqnarray}\label{3r}
    a_0 &=& \frac{(\sin \phi + \cos \phi)\cos \phi}{1+\sin \phi \cos
    \phi}\cr && \cr
    a_1 &=& \frac{(\sin \phi + \cos \phi)\sin \phi}{1+\sin \phi \cos
    \phi}\cr &&\cr
    a_2 &=& -\frac{(\sin \phi \cos \phi)}{1+\sin \phi \cos
    \phi}.
\end{eqnarray}
Thus we find a single parameter family of interpolating basis. For
$\phi=0$, we find $|\psi_{00}\ra=|0,0\ra$ and hence the basis
becomes disentangled. Using equation (\ref{ent}) we can calculate
the entanglement. It is plotted in figure (\ref{E3a}) as a function
of $\phi$.

The maximum entanglement is approximately equal to $0.87$. This
shows that in a real three dimensional space, we can not interpolate
to a maximally entangled basis at least by the construction
presented in this paper. However an interpolating basis exists in a
complex three dimensional space as we show in the next subsection.\\

\subsection{Case b: Complex vector space}
Again in this case equations (\ref{ortho}) reduce to two equations
\begin{eqnarray}\label{100}
    |a_0|^2+|a_1|^2+|a_2|^2&=&1\cr &&\cr
    \overline{a_0}a_1+\overline{a_1}a_2+\overline{a_2}a_0&=&0.
\end{eqnarray}

A solution for this set of equations leads to the following state
$|\psi_{00}\ra$
\begin{equation}\label{}
    |\psi_{00}\ra=\frac{1}{\sqrt{1+8\cos^2 \phi}}(2\cos \phi |0,0\ra - e^{i\phi}|1,1\ra + 2\cos \phi
    |2,2\ra).
\end{equation}
The reader can read the coefficients from the above state and can
check that equations (\ref{100}) are satisfied. For convenience we
list here the nine states of the basis, with
$N:=\frac{1}{\sqrt{1+8\cos^2\phi}}$, they read
\begin{eqnarray}\label{3c}
    |\psi_{00}\ra &=& N(2\cos \phi |0,0\ra - e^{i\phi}|1,1\ra + 2\cos \phi
    |2,2\ra)\cr &&\cr
 |\psi_{01}\ra &=& N(2\cos \phi |0,1\ra - e^{i\phi}|1,2\ra + 2\cos \phi
    |2,0\ra)\cr &&\cr
     |\psi_{02}\ra &=& N(2\cos \phi |0,2\ra - e^{i\phi}|1,0\ra + 2\cos \phi
    |2,1\ra)\cr &&\cr
     |\psi_{10}\ra &=& N(2\cos \phi |1,1\ra - e^{i\phi}|2,2\ra + 2\cos \phi
    |0,0\ra)\cr &&\cr
     |\psi_{11}\ra &=& N(2\cos \phi |1,2\ra - e^{i\phi}|2,0\ra + 2\cos \phi
    |0,1\ra)\cr &&\cr
     |\psi_{12}\ra &=& N(2\cos \phi |1,0\ra - e^{i\phi}|2,1\ra + 2\cos \phi
    |0,2\ra)\cr &&\cr
     |\psi_{20}\ra &=& N(2\cos \phi |2,2\ra - e^{i\phi}|0,0\ra + 2\cos \phi
    |1,1\ra)\cr &&\cr
     |\psi_{21}\ra &=& N(2\cos \phi |2,0\ra - e^{i\phi}|0,1\ra + 2\cos \phi
    |1,2\ra)\cr &&\cr
     |\psi_{22}\ra &=& N(2\cos \phi |2,1\ra - e^{i\phi}|0,2\ra + 2\cos \phi
    |1,0\ra).
\end{eqnarray}

The reader can easily check that this basis is orthogonal.
Moreover this basis interpolates between a product basis for $\phi
= \frac{\pi}{2}$ and a maximally entangled basis for $\phi =
\frac{\pi}{3}$. The entanglement of this basis is plotted as a
function of $\phi$ in
figure (\ref{E3a}).\\


\section{Equi-entangled basis for four dimensional
spaces}\label{four}
\subsection{Case a: Real vector space}
In four dimensions we can find an interpolating basis for real
vector spaces. In this case equations (\ref{ortho}) reduce to the
following three equations:
\begin{eqnarray}\label{3}
a_0^2+a_1^2+a_2^2+a_3^2&=&1\cr a_0a_2+a_1a_3&=&0\cr &&\cr
(a_0+a_2)(a_1+a_3)&=&0.
   \end{eqnarray}
A solution (not the only solution) of these equations is provided by
\begin{equation}\label{4}
    a_0=\frac{1}{2}\cos \theta, \ \ \  a_1 = \frac{1}{2}(1+\sin \theta),
    \ \ \  a_2=\frac{-1}{2}\cos \theta, \ \ \
    a_3=\frac{1}{2}(1-\sin\theta),
    \end{equation}
leading to the state $|\psi_{00}\ra$,

\begin{equation}\label{4r}
    |\psi_{00}\ra= \frac{1}{2}\cos \theta|00\ra+\frac{1}{2}(1+\sin \theta)|11\ra-\frac{1}{2}\cos \theta|22\ra+\frac{1}{2}(1-\sin
    \theta)|33\ra.
\end{equation}

\begin{figure}[t]
 \centering
   \includegraphics[width=12cm,height=8cm]{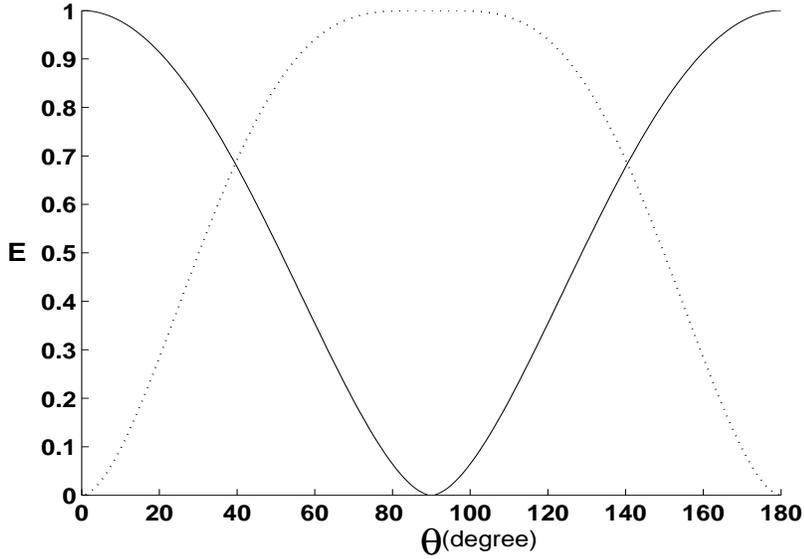}
   \caption{a)The entanglement of the basis (\ref{4r}) as a function of $\theta$ (in degree)(----).
            b)The entanglement of the basis (\ref{4c}) as a function of $\theta$ (in degree)(....). The entanglement is dimensionless.}
   \label{E4a}
\end{figure}

The basis thus constructed is a separable state for
$\theta=\frac{\pi}{2}$ and a maximally entangled for $\theta=0$.
Figure (\ref{E4a}) shows the entanglement of the basis as a
function of $\theta$.

\subsection{Case b: Complex vector space} In this case
equations (\ref{ortho}) reduce to the three independent equations

\begin{eqnarray}\label{3}
|a_0|^2+|a_1|^2+|a_2|^2+|a_3|^2&=&1\cr &&\cr
\overline{a_0}a_1+\overline{a_1}a_2 + \overline{a_2}a_3 +
\overline{a_3} a_0&=&0\cr  &&\cr \overline{a_0}a_2+\overline{a_1}a_3
+ \overline{a_2}a_0 + \overline{a_3} a_1&=&0.
\end{eqnarray}

We present one of the solutions of this set, (for more general
solutions see next section).  Let

\begin{eqnarray}\label{4c}
  &&a_0 = \frac{1}{2}(1+e^{i\theta}\cos \theta)  \cr
  &&a_1=ia_2=-a_3  = \frac{1}{2}e^{i\theta} \sin \theta.
  \end{eqnarray}

The corresponding basis interpolates between a separable basis (for
$\theta = 0 $) where
\begin{equation}
|\psi_{00}(\theta = 0)\ra=|00\ra
\end{equation}
to a maximally entangled basis (for $\theta= \frac{\pi}{2}$), where
\begin{equation}
|\psi_{00}(\theta = \frac{\pi}{2})\ra=\frac{1}{2}(|00\ra +
i|11\ra + |22\ra - i|33\ra).
\end{equation}
A simple calculation shows that the entanglement of this basis is
equal to
\begin{equation}\label{S3}
    E = - \lambda_0 \log_4 \lambda_0 - 3 \lambda_1 \log_4 \lambda_1,
    \end{equation}
where $\lambda_0 = \frac{1}{4}(1+3 \cos^2 \theta)$  and
$\lambda_1 = \frac{1}{4}\sin^2 \theta$. Figure (\ref{E4a}) shows
the value of entanglement as a function of the parameter
$\theta$. It is seen that for a rather broad range of the
parameter, the state is nearly maximally entangled.

Until now we have restricted ourselves to low-dimensional spaces.
In the next section we provide a general solution of $d$
dimensional spaces.
\section{Equi-entangled bases for $d$ dimensional
spaces}\label{higher}
\begin{table}
  \centering
  \begin{tabular}{|c|c|c|c|c|c|c|c|c|c|c|c|}

         \hline
         &  &  &  &  &  &  &  &  &  & \\
         &$\theta_0^0$ & $\theta_1^0$ & $\theta_2^0$ & $\theta_3^0$ & $\theta_4^0$ & $a_0$ & $a_1$ & $a_2$  & $a_3$ & $a_4$\\
         &  &  &  &  &  &  &  &  &  & \\
         \hline
         &  &  &  &  &  &  &  &  &  & \\
     d=2 & 0 & $\frac{\pi}{2}$ & - & - & - & $\frac{1}{\sqrt{2}}$ & $\frac{i}{\sqrt{2}}$ & - & -&-\\
         &  &  &  &  &  &  &  &  &  & \\
         \hline
         &  &  &  &  &  &  &  &  &  & \\
     d=3 & 0 & 0 & $\frac{2\pi}{3}$ & - & - & $\frac{1}{\sqrt{3}}$ & $\frac{1}{\sqrt{3}}$ & $\frac{-i}{\sqrt{3}}$ & -&-\\
         &  &  &  &  &  &  &  &  &  & \\
         \hline
         &  &  &  &  &  &  &  &  &  & \\
     d=4 & 0 & 0 & 0 & $\pi$ & - & $\frac{1}{2}$& $\frac{i}{2}$ & $\frac{1}{2}$ & $\frac{-i}{2}$ &-\\
         &  &  &  &  &  &  &  &  &  & \\
         \hline
         &  &  &  &  &  &  &  &  &  & \\
    d=4  & 0 & $\pi$ & $\pi$ & $\pi$ & - & $\frac{i}{2}$ & $\frac{1}{2}$ & $\frac{1}{2}$ & $\frac{1}{2}$ & -\\
         &  &  &  &  &  &  &  &  &  & \\
         \hline
         &  &  &  &  &  &  &  &  &  & \\
    d=5  & 0 & $\frac{2\pi}{5}$ & 0 & $\frac{4\pi}{5}$ & $\frac{4\pi}{5}$ & $\frac{1}{\sqrt{5}}e^{\frac{2i\pi}{5}}$ & $\frac{1}{\sqrt{5}}e^{\frac{2i\pi}{5}}$& $\frac{1}{\sqrt{5}}e^{-\frac{2i\pi}{5}}$ & $\frac{1}{\sqrt{5}}$ &$\frac{1}{\sqrt{5}}e^{-\frac{2i\pi}{5}}$\\
         &  &  & & & &  &  &  & & \\
         \hline
  \end{tabular}
  \caption{The phases $\theta_{\a}^0$ and the coefficients $a_{k}$'s for two,three, four and five dimensional
spaces.In each case we have used the freedom to multiply all the
coefficients $a_{k}$ by a phase.}\label{table1}
\end{table}
In order to solve equations (\ref{ortho}) for spaces of arbitrary
dimensions we write them in matrix form as follows:
\begin{equation}\label{orthomatrix}
     A^{\dagger} S^m A =\delta_{m,0} \h m=0,1,2,\cdots d-1,
\end{equation}
where $A$ is the vector of coefficients $A:=(a_0,a_1,\cdots
a_{d-1})^T$. In order to solve these later equations we expand the
vector $A$ in terms of the eigenvectors of $S$. The matrix $S$
having the property $S^d=I$, has the following spectrum
\begin{equation}\label{spectrum}
    Sv_{\a}= \xi^{\a}v_{\a},
\end{equation}
where $\xi=e^{\frac{2\pi i}{d}} $ and
\begin{equation}\label{v}
    (v_{\a})_j:= \frac{1}{\sqrt{d}}\xi^{\a j}.
\end{equation}
Inserting the expansion of the vector $A$ in terms of the
eigenvectors of $S$ as

\begin{equation}\label{A}
    A=\sum_{\a=0}^{d-1} c_{\a} v_{\a},
\end{equation}

into (\ref{orthomatrix}), we find a system of linear equations for
the absolute square of the coefficients $|c_{\a}|^2$, given by

\begin{equation}\label{bs}
    \left(\begin{array}{ccccc} 1 & 1 & 1& \cdots  & 1 \\
   1 & \xi & \xi^2 & \cdots  & \xi^{d-1}\\
   1 & \xi^2 & \xi^4 &\cdots  & \xi^{2(d-1)}\\
   \cdot & \cdot & \cdot & \cdots  &\cdot \\
1 & \xi^{d-1} & \xi^{2(d-1)} &\cdots  & \xi^{(d-1)^2}\\
\end{array}\right)
\left(\begin{array}{c}|c_0|^2\\
|c_1|^2 \\ |c_2|^2
\\ \cdot  \\ |c_{d-1}|^2\end{array}\right)=\left(\begin{array}{c}1\\
0 \\ 0  \\ \cdot \\ 0\end{array}\right).
\end{equation}
The unique solution of this last equation is given by
\begin{equation}\label{cc}
    |c_0|^2=|c_1|^2=\cdots |c_{d-1}|^2=\frac{1}{d},
\end{equation}
or
\begin{equation}\label{ccc}
    c_{\a}=\frac{1}{\sqrt{d}}e^{i\theta_{\a}}, \h \a=0,1,\cdots d-1,
\end{equation}
where $\theta_{\a}$ are $d$ free parameters. Of these, one
parameter can be absorbed into the total phase of state and we
remain with $d-1$ continuous parameters.  Inserting these values
in the expansion of $A$,we find the final form of the desired
coefficients,
\begin{equation}\label{afinals}
    a_k:= \frac{1}{d}\sum_{\a=0}^{d-1} e^{i\theta_{\a}}\xi^{k\a}.
\end{equation}
Thus we have found a $d-1$ parameter family of $d^2$ orthonormal
states all of which have the same value of entanglement. When all
the parameters $\theta_{\a}=0$, we find from (\ref{afinals}) that
$a_k = \delta_{k,0}$ and hence $|\psi_{0,0}\ra$ and all the other
states become separable. By varying the parameters $\theta_{\a}$,
we can continuously change the entanglement of the basis. The
question of wether one can actually reach maximally entangled
states in this family amounts to finding solutions for
$\theta_{\a}$ in the system of equations (\ref{afinals}) such
that the left hand sides are all $\frac{1}{\sqrt{d}} \times $pure
phases. This question is difficult to answer in its generality.

We have tried to find solutions in low dimensional spaces, although
we would like to stress that the very existence of such solutions in
arbitrary dimensions is not certain to us. We present some of the
solutions below in table (\ref{table1}). Having such solutions which
we denote by $\theta_{\a}^0$ one can easily define an interpolating
basis by writing $\theta_{\a}=t \theta_{\a}^0$, where $t\in [0,1]$.

\section{Discussion}

We have posed a question on the existence of a basis for a tensor
product of two $d$ dimensional spaces (qudits) so that by changing
some parameters, this basis while being orthonormal, changes
continuously from a product basis into a maximally entangled
basis. We have also required that all the states of such a basis
have the same value of entanglement during the interpolation. We
call such a basis an equi-etangled orthonormal basis. We have
found such bases explicitly in three and four and five dimensional
spaces and have given a general solution for spaces of arbitrary
dimensions. Such bases are of theoretical and possibly
experimental interest for several reasons. First they provide us
with a generalized Bell state measurement in which the output
states are not necessarily maximally entangled but states with a
prescribed value of entanglement. This application may be
interesting in view of the recent results that in dimensions
higher than two, maximally entangled states are not necessarily
the states which have maximal non-locality.  Second they provide
us with a tool for analyzing the effect of entanglement in the
capacity of quantum channels, since we can encode the input
alphabet into mutually distinguishable entangled codewords.
Finally they may be of use in quantum cryptography with $d$ level
states.

\section{Acknowledgements}
We would like to thank M. Asoudeh, A. Bayat, F. Ghahhari, I.
Marvian, and M. Shahrokhshai for valuable discussions.

{}
\end{document}